\documentclass[a4paper,10pt,twoside]{article}

\usepackage{amssymb}
\usepackage{amsmath}
\usepackage{graphics}

\usepackage{amssymb,amsmath,graphicx}

\makeatletter
\let\@afterindentfalse\@afterindenttrue
\@afterindenttrue
\makeatother

\newtheorem{definition}{Definition}
\newtheorem{theorem}{Theorem}

\newtheorem{corollary}{Corollary}

\newcommand{\vt}{\vartheta}
\newcommand{\la}{\lambda}
\newcommand{\Pn}{\mathcal{P}_{n}}
\newcommand{\Ptwo}{\mathcal{P}_{2}}
\newcommand{\Pthree}{\mathcal{P}_{3}}
\newcommand{\Ta}{\mathop{\mathrm{T}}\nolimits}
\newcommand{\Tr}{\mathop{\mathrm{Tr}}\nolimits}
\newcommand{\ci}{\mathop{\mathrm{i}}\nolimits}
\newcommand{\sa}{\mathop{\mathrm{sa}}\nolimits}
\newcommand{\Scal}{\mathop{\mathrm{Scal}}\nolimits}
\newcommand{\KM}{\mathop{\mathrm{KM}}\nolimits}
\newcommand{\dint}{\mathop{\mathrm{d}}\nolimits}
\newcommand{\id}{\mathop{\mathrm{id}}\nolimits}
\newcommand{\MN}{\mathcal{M}_{n}}
\newcommand{\Mtwo}{\mathcal{M}_{2}}
\newcommand{\btop}[2]{\genfrac{}{}{0pt}{2}{#1}{#2}}

\begin{document}

\title{
  On the curvature of the quantum state space with pull-back metrics
  \thanks{keywords: state space, monotone statistical metric, scalar curvature;
          MSC: 53C20, 81Q99}}
\author{Attila Andai\thanks{andaia@math.bme.hu}\\
  Department for Mathematical Analysis,\\
  Budapest University of Technology and Economics,\\
  H-1521 Budapest XI. Sztoczek u. 2, Hungary}
\date{April 6, 2006}

\maketitle

\begin{abstract}
The aim of the paper is to extend the notion of $\alpha$-geometry in the classical and in the noncommutative case
  by introducing a more general class of pull-back metrics and to give concrete formulas for the scalar curvature
  of these Riemannian manifolds.
We introduce a more general class of pull-back metrics of the noncommutative state spaces, we pull back the
  Euclidean Riemannian metric of the space of self-adjoint matrices with functions which have an analytic extension
  to a neighborhood of the interval $\left]0,1\right[$ and whose derivative are nowhere zero.
We compute the scalar curvature in this setting, and as a corollary we have the scalar curvature
  of the classical probability space when it is endowed with such a general pull-back metric.
In the noncommutative setting we consider real and complex state spaces too.
We give a simplification of Gibilisco's and Isola's conjecture for the first nontrivial classical probability
  space and we present the result of a numerical computation which indicate that the conjecture may be true
  for the space of real and complex qubits.
\end{abstract}

\section*{Introduction}

The idea that the space of probability distributions can be endowed with Riemannian metric
  is due to Rao \cite{Rao}, and it was developed by Cencov \cite{Cen},
  Amari and Nagaoka \cite{Ama,AmaNag} and Streater \cite{Str} among others.
Cencov and Morozova \cite{CenMor} were the first to study the monotone metrics on classical statistical manifolds.
They proved that such a metric is unique, up to normalization.
The counterpart of this theorem in quantum setting was given by Petz \cite{Pet1}, who showed that
  monotone metrics can be labeled by special operator monotone functions.
Some differential geometrical quantities were computed for these manifolds with monotone
  metrics, one of them is the scalar curvature \cite{Dit1,Dit2,MPA,PetSud}.
The scalar curvature at every state measures the average statistical uncertainty of the state \cite{Pet3,Pet2}.
This is one of the basic ideas of Petz's conjecture \cite{Pet2}, which is about the monotonicity of the
  scalar curvature with respect to the majorization relation when the state space is endowed with the
  Kubo--Mori metric.
This conjecture is still unsolved, partial results can be found in \cite{And1,And2,Dit1,GibIso1,HPT,Pet2}.

Cencov introduced the $\alpha$-connections and $\alpha$-geometry on the space of classical probability
  distributions \cite{Cen}, it was developed by Amari and Nagaoka \cite{AmaNag},
  Gibilisco and Pistone \cite{GibPis}.
Gibilisco and Isola showed that the idea of Petz's conjecture can be extended to $\alpha$-geometries.
They also have another conjecture about the monotonicity of the scalar curvature when the classical and
  the noncommutative probability spaces are endowed with $\alpha$-geometry \cite{GibIso1}.
This conjecture was proved for the space of probability distributions on a set which consists of two elements.
They used the classical curvature formula for this one dimensional manifold, since its scalar curvature is zero.

The aim of the paper is to extend the notion of $\alpha$-geometry in the classical and in the noncommutative case
  by introducing a more general class of pull-back metrics and to give concrete formulas for the scalar curvature
  of these Riemannian manifolds.
The classical and noncommutative probability spaces are one codimensional submanifolds of flat spaces.
We use a general formula from differential geometry to compute the curvature of these one codimensional
  submanifolds.
In the first section we do this computation for classical probability spaces when they are endowed with
  a special pull-back metric, with $\alpha$-geometry.
In the second section we introduce a more general class of pull-back metrics of the noncommutative state
  spaces, we pull back the Euclidean Riemannian metric of the space of self-adjoint matrices with
  functions which have analytic extension to a neighborhood of the interval $\left]0,1\right[$ and
  whose derivative are nowhere zero.
We compute the scalar curvature in this setting, and as a corollary we have the scalar curvature
  of the classical probability space when it is endowed with such a general pull-back metric.
In the noncommutative setting we consider real and complex state spaces too.
We check the theorems in some special cases when the result is known from somewhere else,
  since the computation is a bit lengthy.
Finally in the third section we give a simplification of the conjecture for the first nontrivial
  classical probability space and we present the result of a numerical computation which indicate
  that the conjecture may be true for the space of real and complex qubits.

\section{Classical $\alpha$-geometries}

We work on a special classical statistical manifold, on the space of probability distributions
  on a finite set.

\begin{definition}
For every number $n\in\mathbb{N}^{+}$ let $\Pn$ denote the open set of the probability distributions on a
  space which consists of $n$ points, that is
\begin{equation*}
\Pn=\left\lbrace (\vt_{1},\dots,\vt_{n})\in\mathbb{R}^{n}\ \Bigm\vert\
\forall k\in\{1,\dots,n \}:\vt_{k}>0,\ \sum_{k=1}^{n}\vt_{k}=1\right\rbrace.
\end{equation*}
\end{definition}

The majorization relation is one of the most important relation between probability distributions.

\begin{definition}
The distribution $a=(a_{1},\dots,a_{n})\in\Pn$ is said to be majorized by the distribution
  $b=(b_{1},\dots,b_{n})\in\Pn$, denoted by $a\prec b$ if the following inequalities hold for their
  decreasingly ordered set of parameters $(a_{i}^{\downarrow})_{i=1,\dots,n}$ and
  $(b_{i}^{\downarrow})_{i=1,\dots,n}$
\begin{equation*}
\sum_{l=1}^{k}a_{l}^{\downarrow}\leq \sum_{l=1}^{k}b_{l}^{\downarrow}
\end{equation*}
for all $1\leq k<n$.
\end{definition}

The intuitive meaning of the majorization relation $a\prec b$ is that the distribution $a$ is more mixed or
  more chaotic than the distribution $b$.

The tangent space of the Riemannian manifold $(M,g)$ at a point $p\in M$ will be denoted by $\Ta_{p}M$,
  and the tangent bundle $\displaystyle\bigcup_{p\in M}\Ta_{p}M$ will be denoted by $\Ta M$.
The canonical Riemannian metric $g_{c}$ of the spaces $M=\mathbb{R}^{n},\mathbb{R}_{+}^{n}$
  at every point $p\in M$ for every tangent vectors $x,y\in\Ta_{p}M$ is
\begin{equation*}
g_{c}(p)(x,y)=\sum_{i=1}^{n}x_{i}y_{i}.
\end{equation*}
The space $\Pn$ is a differentiable manifold and one can endow it with Riemannian metric in many ways,
  one family of these metrics is called $\alpha$-geometry.

\begin{definition}
For every parameter $\alpha\in\mathbb{R}$ the $\alpha$-geometry of $\Pn$ is the pull-back
  geometry of the Riemannian manifold $(\mathbb{R}^{n},g_{c})$ induced by the map
\begin{equation*}
\phi_{\alpha,n}:\Pn\to\mathbb{R}^{n}\qquad(\vt_{1},\dots,\vt_{n})\mapsto
  \left\lbrace \begin{array}{lll}
  \frac{2}{1-\alpha}\left(\vt_{1}^{\frac{1-\alpha}{2}},\dots,\vt_{n}^{\frac{1-\alpha}{2}}\right),
                                   & \mbox{if} & \alpha\neq 1\\[1em]
  (\log\vt_{1},\dots,\log\vt_{n}), & \mbox{if} & \alpha=1. \end{array}\right.
\end{equation*}
This Riemannian space is denoted by $(\Pn,g_{\alpha})$.
\end{definition}

For further computations it will be useful to extend this Riemannian space to non-normalized distributions.

\begin{definition}
The extended Riemannian space of $(\Pn,g_{\alpha})$ is $(\Tilde{\Pn},\tilde g_{\alpha})$, where
  $\Tilde{\Pn}=\mathbb{R}_{+}^{n}$ and the Riemannian metric $\tilde g_{\alpha}$ is the pull-back
  geometry of $(\mathbb{R}_{+}^{n},g_{c})$ metric induced by the map
\begin{equation*}
\tilde\phi_{\alpha,n}:\Tilde\Pn\to\mathbb{R}^{n}\qquad(\vt_{1},\dots,\vt_{n})\mapsto
  \left\lbrace \begin{array}{lll}
  \frac{2}{1-\alpha}\left(\vt_{1}^{\frac{1-\alpha}{2}},\dots,\vt_{n}^{\frac{1-\alpha}{2}}\right),
                                   & \mbox{if} & \alpha\neq 1\\[1em]
  (\log\vt_{1},\dots,\log\vt_{n}), & \mbox{if} & \alpha=1. \end{array}\right.
\end{equation*}
\end{definition}

To simplify the notations we fix the parameter $\alpha\neq 1$, we denote the $\alpha$-Riemannian metric
  with $g$, $\tilde{g}$ and we set $\beta=-\alpha-1$.
The space $(\Pn,g)$ is a subspace of $(\Tilde{\Pn},\tilde{g})$, and the metric $g$ is the pull-back
  of $\tilde{g}$ induced by the natural $i:\Pn\to\Tilde{\Pn}$ embedding.
At a point $\vt=(\vt_{1},\dots,\vt_{n})\in\Tilde{\Pn}$ the tangent space $\Ta_{\vt}\Tilde{\Pn}$ consists
  of the vectors of the form $a_{1}\partial_{1}+\dots+a_{n}\partial_{n}$ and the Riemannian metric is
  $\tilde{g}_{ij}=\tilde{g}(\partial_{i},\partial_{j})=\delta_{ij}\vt_{i}^{\beta}$.
If $\vt\in\Pn$ then $\Ta_{\vt}\Pn$ consists of those vectors  of the form
  $a_{1}\partial_{1}+\dots+a_{n}\partial_{n}$ of $\Ta_{\vt}\Tilde{\Pn}$
  for which $\displaystyle\sum_{k=1}^{n}a_{k}=0$.
The equality $g_{ij}=\tilde{g}_{ij}$ holds for the metrics on the subspace $\Ta_{\vt}\Pn$.

To compute the scalar curvature of the space $\Pn$ it is convenient to consider it as a submanifold of
$\Tilde{\Pn}$ and to use the following theorem from differential geometry.

\begin{theorem}
\label{th:scal main theorem}
Assume that $(M,g)$ is an $n$ dimensional submanifold of the $n+1$ dimensional Riemannian space
  $(\Tilde{M},\tilde{g})$, such that $g$ is the pull-back metric induced by the natural embedding $M\to\Tilde{M}$.
The Levi--Civita covariant derivative of $\Tilde{M}$ is denoted by $\Tilde{\nabla}$ and the Riemannian
  curvature tensor of $\Tilde{M}$ is denoted by $\Tilde{R}$.
The normal vector field of $M$ is $N:M\to\Ta\Tilde{M}$.
For every tangent vector $X,Y\in\Ta\Tilde{M}$ let us define the following map.
\begin{equation}\label{eq:S definition}
S(X,Y):M\to\mathbb{R}\qquad p\mapsto -\tilde{g}(\Tilde{\nabla}_{X}N,Y)
\end{equation}
For every point $p\in M$ if $(A_{t})_{t=1,\dots,n}$ is an orthonormal basis in $\Ta_{p}M$
  (that is $g(A_{t},A_{s})=\delta_{ts}$) then the scalar curvature of $M$ at a point $p$ is given by the
  equation
\begin{equation}\label{eq:scal main equation}
\Scal(p)=\sum_{t,s=1}^{n}\tilde{g}(\Tilde{R}(A_{t},A_{s})A_{s},A_{t})
  +S(A_{s},A_{s})S(A_{t},A_{t})-S(A_{t},A_{s})S(A_{s},A_{t}).
\end{equation}
\end{theorem}

The correspondence between the Levi--Civita covariant derivative $\nabla_{\partial_{i}}\partial_{j}$ and
  Christoffel symbol of the second kind $\Gamma_{ij}^{k}$ is given by the equation
  $\displaystyle\nabla_{\partial_{i}}\partial_{j}=\sum_{k=1}^{n}\Gamma_{ij}^{..k}\partial_{k}$.
This symbol can be computed using the derivatives of the metric tensor $\tilde{g}_{ij}$ and its inverse
  $\tilde{g}^{ij}$.
The inverse of the metric tensor is $\tilde{g}^{ij}=\delta_{ij}\vt_{i}^{-\beta}$.
The Christoffel symbol of the second kind of the space $\Tilde{\Pn}$ is the following.
\begin{align}\label{eq:notimportant2}
 \Tilde{\Gamma}_{ij}^{..k}&=\frac{1}{2}\sum_{m=1}^{n}
   \tilde{g}^{km}(\partial_{i}\tilde{g}_{jm}+\partial_{j}\tilde{g}_{im}-\partial_{m}\tilde{g}_{ij} )\\[0.5em]
  &=\frac{1}{2}\sum_{m=1}^{n}\delta_{km}\vt_{m}^{-\beta}(\delta_{jm}\delta_{ik}\beta\vt_{i}^{\beta-1})=
  \frac{\beta}{2}\vt_{k}^{-1}\delta_{ij}\delta_{jm}\nonumber
\end{align}
The space $\Tilde{\Pn}$ is diffeomorphic to $\mathbb{R}^{n}$ so for the curvature tensor
  $\Tilde{R}_{ijk}^{...i}=0$ holds.
The normal vector field of the submanifold $\Pn$ is
\begin{equation*}
N(\vt)=\frac{1}{c(\vt)}\sum_{i=1}^{n}\vt_{i}^{-\beta}\partial_{i}, \quad\mbox{where}\quad
  c(\vt)=\sqrt{\sum_{i=1}^{n}\vt_{i}^{-\beta}}\ ,
\end{equation*}
since
\begin{align*}
&\tilde{g}(N,N)=\frac{1}{c(\vt)^{2}}\sum_{i=1}^{n}\vt_{i}^{-2\beta}\vt_{i}^{\beta}=1,\quad\mbox{and} \\
&\tilde{g}(N,\partial_{i}-\partial_{n})=\frac{1}{c(\vt)}
  (\tilde{g}(\vt_{i}^{-\beta}\partial_{i},\partial_{i})-\tilde{g}(\vt_{n}^{-\beta}\partial_{n},\partial_{n}))
  =0.
\end{align*}

\vfill\eject\newpage

The covariant derivative of the normal vector field at a the point $\vt\in\Pn$ is
\begin{align*}
(\Tilde{\nabla}_{\partial_{i}}N)(\vt)&=\sum_{k=1}^{n}\Tilde{\nabla}_{\partial_{i}}
  \left(\frac{1}{c(\vt)}\vt_{k}^{-\beta}\partial_{k} \right)=
  \sum_{k=1}^{n}\partial_{i}\left(\frac{1}{c(\vt)}\vt_{k}^{-\beta} \right)\partial_{k}
  +\frac{1}{c(\vt)}\vt_{k}^{-\beta}\Tilde{\nabla}_{\partial_{i}}\partial_{k}\\
&= \sum_{k=1}^{n}\left(\frac{\beta}{c(\vt)^{3}}\vt_{i}^{-\beta-1}\vt_{k}^{-\beta}
  -\frac{\beta}{c(\vt)}\delta_{ik}\vt_{i}^{-\beta-1} \right)\partial_{k}
  +\frac{\beta}{2c(\vt)}\delta_{ik}\vt_{k}^{-\beta-1}\partial_{k}\\
&= -\frac{\beta}{2c(\vt)}\vt_{i}^{-\beta-1}\partial_{i}
  +\frac{\beta}{2c(\vt)^{3}}\vt_{i}^{-\beta-1}\sum_{k=1}^{n}\vt_{k}^{-\beta}\partial_{k}.
\end{align*}
At the third equality in this computation we used the definition of the Christoffel symbol and
  Equation (\ref{eq:notimportant2}).
For the tangent vectors $\partial_{i},\partial_{j}$ the value of the function $S$ is the following.
\begin{equation}\label{eq:notimportant3}
S(\partial_{i},\partial_{j})=-\tilde{g}(\Tilde{\nabla}_{\partial_{i}}N,\partial_{j})=
  \frac{\beta}{2c(\vt)}\delta_{ij}\vt_{i}^{-1}-\frac{\beta}{c(\vt)^3}\vt_{i}^{-\beta-1}.
\end{equation}

To compute the scalar curvature of the space $\Pn$ we use Equation (\ref{eq:scal main equation}).
For every point $\vt\in\Pn$ if $(A_{t})_{t=1,\dots,n-1}$ is an orthonormal basis in $\Ta_{\vt}\Pn$,
  $(A_{t})_{t=1,\dots,n-1}\cup N(\vt)$ is an orthonormal basis in $\Ta_{\vt}\Tilde{\Pn}$.
Since the space $(\Tilde{\Pn},\tilde{g})$ is flat and in the Equation (\ref{eq:scal main equation})
  the sum is independent from the orthonormal basis, we can compute the scalar curvature as
\begin{align}\label{eq:notimportant1}
\Scal(\vt)=&\sum_{t,s=1}^{n} \bigl(S(B_{s},B_{s})S(B_{t},B_{t})-S(B_{t},B_{s})S(B_{s},B_{t})\bigr)\\
        &-2\sum_{t=1}^{n} \bigl(S(N(\vt),N(\vt))S(B_{t},B_{t})-S(B_{t},N(\vt))S(N(\vt),B_{t})\bigr),\nonumber
\end{align}
  where $(B_{t})_{t=1,\dots,n}$ is an orthonormal basis in $\Ta_{\vt}\Tilde{\Pn}$.
For every $\vt\in\Pn$ we have
\begin{align*}
&S(\partial_{i},N(\vt))=\frac{1}{c(\vt)}\sum_{k=1}^{n}\vt_{k}^{-\beta}S(\partial_{i},\partial_{k})
 =\frac{\beta\vt_{i}^{-\beta-1}}{2c(\vt)^2}-
 \frac{\beta}{2c(\vt)^4}\sum_{k=1}^{n}\vt_{k}^{-\beta}\vt_{i}^{-\beta-1}=0\\
&S(N(\vt),N(\vt))=\frac{1}{c(\vt)}\sum_{k=1}^{n}\vt_{k}^{-\beta}S(\partial_{k},N(\vt))=0,
\end{align*}
  it means that in Equation (\ref{eq:notimportant1}) the second summation  is $0$.
The set of vectors $\left(\vt_{t}^{-\frac{\beta}{2}}\partial_{t} \right)_{t=1,\dots,n}$ form an orthonormal basis
  in $\Ta_{\vt}\Tilde{\Pn}$, therefore
\begin{equation*}
\Scal(\vt)=\sum_{t,s=1}^{n}
  \vt_{s}^{-\beta}S(\partial_{s},\partial_{s})\vt_{t}^{-\beta}S(\partial_{t},\partial_{t})-
  \vt_{s}^{-\beta}\vt_{t}^{-\beta}S(\partial_{t},\partial_{s})S(\partial_{s},\partial_{t}).
\end{equation*}
Substituting Equation (\ref{eq:notimportant3}) into the previous one after simplification we have the
  following theorem.

\begin{theorem}\label{th:classical scalar curvature}
The scalar curvature of the space $(\Pn,g_{\alpha})$ at a point $\vt\in\Pn$ is
\begin{equation}\label{eq:scal classical equation}
\Scal(\vt)=\frac{(1+\alpha)^{2}}{4c(\vt)^2}\sum_{\btop{t,s=1}{t\neq s}}^{n}\vt_{t}^{\alpha}\vt_{s}^{\alpha}
 \left(1-\frac{\vt_{t}^{\alpha+1}+\vt_{s}^{\alpha+1}}{c(\vt)^2} \right),
\end{equation}
where $\displaystyle c(\vt)=\sqrt{\sum_{k=1}^{n}\vt_{k}^{\alpha+1}}$ .
\end{theorem}

We proved this Theorem now just for the $\alpha\neq 1$ case, but in the next section we show that
 this is true for the $\alpha=1$ case too, see Equation (\ref{eq:scal classical equation log}).

One can check the previous theorem in two special cases easily.
The parameter $\alpha=-1$ corresponds to the case, when $\phi_{\alpha,n}:\Pn\to\mathbb{R}^{n}$
  is the natural embedding.
In this case $\Pn$ is a part of an $n-1$ dimensional hyperplane, so its scalar curvature is zero.
At the parameter value $\alpha=0$ the function $\phi_{\alpha,n}$ maps $\Pn$ to the surface of the
  Euclidean ball with radius $R=2$.
In this case $c(\vt)=1$ and the scalar curvature formula gives
\begin{align*}
\Scal(\vt)&=\frac{1}{4}\left(\sum_{t,s=1}^n(1-\vt_{t}-\vt_{s})-\sum_{t=1}^{n}(1-2\vt_{t}) \right)=
   \frac{(n-1)(n-2)}{4}\\
 &=\frac{\dim(\Pn)(\dim(\Pn)-1)}{R^{2}}
\end{align*}
which is just the scalar curvature of the $\dim(\Pn)$ dimensional sphere with radius $R$.
For arbitrary $\alpha$ if $n=2$ then the scalar curvature formula gives $0$, since the scalar
  curvature of a one dimensional manifold is $0$.

Finally we note that scalar curvature of the space $(\Pthree,g_{\alpha})$ is a simple formula,
 which is an easy consequence of Theorem (\ref{th:classical scalar curvature}).

\begin{corollary}\label{co:scalar curvature of P3}
The scalar curvature of the space $(\Pthree,g_{\alpha})$ at $\vt=(\vt_{1},\vt_{2},\vt_{3})\in\Pthree$ is
\begin{equation*}
\Scal(\vt)=\frac{(1+\alpha)^{2}}{2}\frac{\vt_{1}^{\alpha}\vt_{2}^{\alpha}\vt_{3}^{\alpha}}
                                        {\left(\vt_{1}^{\alpha+1}+\vt_{2}^{\alpha+1}+\vt_{3}^{\alpha+1} \right)^{2}}.
\end{equation*}
\end{corollary}

\section{Pull-back geometry of the state space}

The quantum mechanical state space on a finite dimensional Hilbert-space is the set of real or complex,
  self adjoint, positive matrices with trace 1.
To put differential geometrical structure to the state space is simpler if we consider the open set
  of positive definite states.

\begin{definition}
For every $n$ let us denote by $\MN$ the set of positive states, that is
\begin{equation*}
\MN=\left\{ D\in M_{n}\ \vert\ D=D^{*},\ D>0,\ \Tr D=1 \right\}.
\end{equation*}
\end{definition}

Some concepts of the classical probability theory can be extended to the noncommutative case.
One of them is the majorization relation.

\begin{definition}
The state $D_{1}\in\MN$ is said to be majorized by the state $D_{2}\in\MN$, denoted by
  $D_{1}\prec D_{2}$, if the relation $\mu_{1}\prec\mu_{2}$ holds for their set of eigenvalues
  $\mu_{1}$ and $\mu_{2}$.
\end{definition}

In the classical case we have defined only a special kind of pull-back metrics, in that case the function
  was a power function or a logarithmic one.
In this quantum setting we consider those $f:\left\rbrack 0,1\right\lbrack\to\mathbb{R}$ functions,
  which have an analytic extension to a neighborhood of the interval $\left\rbrack 0,1\right\lbrack$
  and $f'(x)\neq 0$ for every $x\in\left\rbrack 0,1\right\lbrack$.
We call such functions \emph{admissible functions}.
The set of real or complex self-adjoint matrices will be denoted by $\MN^{\sa}$, and geometrically it will
  be considered as a Riemannian space $(\mathbb{R}^{d},g_{E})$, where $d_{\mathbb{R}}=\frac{(n-1)(n+2)}{2}$
  for real matrices and $d_{\mathbb{C}}=n^{2}-1$ for complex ones and $g_{E}$ is the canonical Riemannian
  metric on $\MN^{\sa}$.
That is, at every point $D\in\MN^{\sa}$ for every vectors $X,Y\in\MN^{\sa}$ in the tangent space at $D$
  the metric is
\begin{equation*}
g_{E}(D)(X,Y)=\Tr XY.
\end{equation*}

\begin{definition}
Assume that $f:\left\rbrack 0,1\right\lbrack\to\mathbb{R}$ is an admissible function.
The pull back of the Riemannian metric $(\MN^{\sa},g_{E})$ to the space $\MN$ induced by the map
\begin{equation*}
\phi_{f,n}:\MN\to\MN^{\sa}\qquad D\mapsto f(D).
\end{equation*}
  is called the pull-back geometry of $\MN$ and it is denoted by $g_{f}$.
This Riemannian space will be denoted by $(\MN,g_{f})$.
\end{definition}

The functions $p\root{p}\of{x}$ if $p\neq 0$ and $\log x$ give back the $\alpha$-geometries.
Assume that there is a given function $f:\left\rbrack 0,1\right\lbrack\to\mathbb{R}$ which is twice
  continuously differentiable and $f'(x)\neq 0$ for every $x\in \left\rbrack 0,1\right\lbrack$.
We will denote the pull-back geometry with $(\MN,g)$, this convention will not cause confusion since
  we fix the function $f$ for further computations.
The computation of the scalar curvature is based on Theorem (\ref{th:scal main theorem}) as in the
  classical case.
If we restrict ourselves to the space of diagonal matrices and to the power or logarithmic functions
  then we get back the formulas of the previous section.

Since the function $f$ has an analytic extension to a neighborhood of the interval $\left\rbrack 0,1 \right\lbrack$
  we have by the Riesz--Dunford operator calculus \cite{Con} for every $D\in\MN$
\begin{equation*}
f(D)=\frac{1}{2\pi\ci}\underset{\gamma}{\oint} f(z)(z\id-D)^{-1}\dint z,
\end{equation*}
  where $\id$ denotes the identity matrix and $\gamma$ is a smooth curve winding once around the spectrum
  of $D$ counterclockwise.
The derivative of $f$ at $D\in\MN$ for $X\in\Ta_{D}\MN$ is
\begin{equation*}
df(D)(X)=\frac{1}{2\pi\ci}\underset{\gamma}{\oint} f(z)(z\id-D)^{-1}X(z\id-D)^{-1}\dint z.
\end{equation*}

Let $D\in\MN$ and choose a basis of $\mathbb{R}^{n}$ such that
  $\displaystyle D=\sum_{i=1}^{n}\la_{i}E_{ii}$ is diagonal, where $(E_{ij})_{1\leq i,j\leq n}$ is the usual
  system of matrix units.
Let us define the following self-adjoint matrices.
\begin{align*}
&F_{ij}=E_{ij}+E_{ji},          &&1\leq i\leq j\leq n;\\
&H_{ij}=\ci E_{ij}-\ci E_{ji},  &&1\leq i< j\leq n.
\end{align*}
The set of matrices $(F_{ij})_{1\leq i\leq j\leq n}\cup (H_{ij})_{1\leq i<j\leq n}$ form a basis of
  $\Ta_{D}\Tilde{\MN}$ for complex matrices and $(F_{ij})_{1\leq i\leq j\leq n}$ form a basis for real ones.

Using the equation
\begin{equation*}
g(D)(X,Y)=\Tr(df(D)(X) df(D)(Y))
\end{equation*}
for the pull-back metric we have the following theorem.

\begin{theorem}
On the Riemannian space $(\MN,g_{f})$ for a state $D\in\MN$ choose a basis of $\mathbb{R}^{n}$ where
  $\displaystyle D=\sum_{i=1}^{n}\la_{i}E_{ii}$.
Then we have for the metric
\begin{align*}
&\mbox{if} &&1\leq i<j\leq n, 1\leq k<l\leq n: \quad
  &&\left\lbrace\begin{array}{l} g(D)(H_{ij},H_{kl})=\delta_{ik}\delta_{jl}2M_{ij}^{2}\\
                               g(D)(F_{ij},F_{kl})=\delta_{ik}\delta_{jl}2M_{ij}^{2}\\
                               g(D)(H_{ij},F_{kl})=0,\end{array}\right.\\
&\mbox{if} &&1\leq i<j\leq n, 1\leq k\leq n: \quad &&g(D)(H_{ij},F_{kk})=g(D)(F_{ij},F_{kk})=0,\\
&\mbox{if} &&1\leq i\leq n,   1\leq k\leq n: \quad &&g(D)(F_{ii},F_{kk})=\delta_{ik}4M_{ii}^{2},
\end{align*}
where
\begin{equation*}
M_{ij}=\left\lbrace\begin{array}{ll}
  \dfrac{f(\la_{i})-f(\la_{j})}{\la_{i}-\la_{j}} &\quad\mbox{if}\ \la_{i}\neq\la_{j}\\
  f'(\la_{i}) &\quad\mbox{if}\ \la_{i}=\la_{j}.\end{array}\right.
\end{equation*}
\end{theorem}

The Christoffel symbol can be computed from the derivative of the Riemannian metric
\begin{equation*}
g(D)(\Gamma(D)(X,Y),Z)=\frac{1}{2}(dg(D)(X)(Y,Z)+dg(D)(Y)(X,Z)-dg(D)(X,Y)).
\end{equation*}
Since the derivative of the Riemannian metric is
\begin{equation*}
dg(D)(Z)(X,Y)=\Tr(d^{2}f(D)(Z)(X)df(D)(Y)+df(D)(X)d^{2}f(D)(Z)(Y))
\end{equation*}
we have the following expression for the Christoffel symbol
\begin{equation*}
\Gamma(D)(X,Y)=(df(D))^{-1}(d^{2}f(D)(X,Y)).
\end{equation*}
From the Riesz--Dunford operator calculus the second derivative of the matrix-valued function $f$ is
\begin{equation*}
d^{2}f(D)(E_{ij})(E_{kl})=\delta_{jk}M_{ilj}E_{il}+\delta_{il}M_{kji}E_{kj},
\end{equation*}
where
\begin{equation*}
M_{ijk}=\frac{1}{2\pi\ci}\underset{\gamma}{\oint}\frac{f(z)}{(z-\la_{i})(z-\la_{j})(z-\la_{k})}\dint z.
\end{equation*}
Combining these results together the Christoffel symbol is the following.
\begin{align}\label{eq:Gamma(D)(F)(F)}
&\Gamma(D)(F_{ij})(F_{kl})=F_{il}\delta_{jk}\frac{M_{ilk}}{M_{il}}+F_{kj}\delta_{il}\frac{M_{ijk}}{M_{jk}}+
                           F_{ik}\delta_{jl}\frac{M_{ikl}}{M_{ik}}+F_{lj}\delta_{ik}\frac{M_{ijl}}{M_{lj}}\\
&\Gamma(D)(H_{ij})(H_{kl})=-F_{il}\delta_{jk}\frac{M_{ilk}}{M_{il}}-F_{kj}\delta_{il}\frac{M_{ijk}}{M_{jk}}+
                           F_{ik}\delta_{jl}\frac{M_{ikl}}{M_{ik}}+F_{lj}\delta_{ik}\frac{M_{ijl}}{M_{lj}}
                           \nonumber\\
&\Gamma(D)(H_{ij})(F_{kl})=H_{il}\delta_{jk}\frac{M_{ilk}}{M_{il}}+H_{kj}\delta_{il}\frac{M_{ijk}}{M_{jk}}+
                           H_{ik}\delta_{jl}\frac{M_{ikl}}{M_{ik}}+H_{lj}\delta_{ik}\frac{M_{ijl}}{M_{lj}}
                           \nonumber
\end{align}

The normal vector field of the submanifold $\MN$ is
\begin{equation}\label{eq:N(D) normal vectorfield}
N(D)=\frac{1}{c(D)}(f'(D))^{-2}, \quad\mbox{where}\quad c(D)=\sqrt{\Tr (f'(D))^{-2}}\,
\end{equation}
since
\begin{align*}
&g(D)(N(D),N(D))=\frac{1}{c(D)^{2}}\sum_{i,j=1}^{n}g(D)
    \left(\frac{1}{M_{ii}^{2}}E_{ii},\frac{1}{M_{jj}^{2}}E_{jj} \right)=1 \\
&g(D)(N(D),E_{ii}-E_{nn})=\frac{1}{c(D)}
  g(D)\left(\frac{1}{M_{ii}^{2}}E_{ii}-\frac{1}{M_{nn}^{2}}E_{nn},E_{ii}-E_{nn} \right)=0.
\end{align*}

In this setting the definition of the map $S$, see Equation (\ref{eq:S definition}), is
\begin{equation}\label{eq:notimportant4}
S(D)(X,Y)=-g(D)(\Gamma(D)(X)(N),Y)
\end{equation}
and now we compute this function.
First we note that
\begin{equation}\label{eq:Gamma(D)(X)(N)}
\Gamma(D)(X)(N)=dN(D)(X)+\Gamma(D)(X)(N(D)).
\end{equation}
Using the $h=(f')^{-2}$ notation the normal vector field is
\begin{equation*}
N(D)=\frac{1}{\sqrt{\Tr h(D)}}h(D)
\end{equation*}
and its derivative is
\begin{equation*}
dN(D)(X)=-\frac{1}{2}\frac{1}{\left(\Tr h(D) \right)^{\frac{3}{2}}}\Tr(dh(D)(X))h(D)
  +\frac{1}{\sqrt{\Tr h(D)}} dh(D)(X).
\end{equation*}
With the function $\phi(D)=D^{-2}$ we have $h=\phi\circ f'$ so the derivative of $h$ is
\begin{equation*}
dh(D)(X)=d\phi(f'(D))(df'(D)(X)).
\end{equation*}
From the Riesz-Dunford operator calculus we have
\begin{equation*}
f'(D)=\frac{1}{2\pi\ci}\underset{\gamma}{\oint}f(z)(z-D)^{-2}\dint z
\end{equation*}
and the derivative of this function is
\begin{equation*}
df'(D)(X)=\frac{1}{2\pi\ci}\underset{\gamma}{\oint}
  f(z)\left((z-D)^{-2}X(z-D)^{-1}+(z-D)^{-1}X(z-D)^{-2}\right)\dint z.
\end{equation*}
It can be explicitly evaluated on the matrix elements
\begin{equation*}
df'(D)(F_{ij})=(M_{iij}+M_{ijj})F_{ij},\quad df'(D)(H_{ij})=(M_{iij}+M_{ijj})H_{ij}.
\end{equation*}
The derivative of the function $\phi$ is
\begin{equation*}
d\phi(D)(X)=-D^{-2}XD^{-1}-D^{-1}XD^{-2}
\end{equation*}
so the derivative of the function $h$ is
\begin{align*}
dh(D)(E_{ij})=&-(f'(D))^{-2}(M_{iij}+M_{ijj})E_{ij}(f'(D))^{-1}\\
              &-(f'(D))^{-1}(M_{iij}+M_{ijj})E_{ij}(f'(D))^{-2}\\
&=-\frac{(M_{iij}+M_{ijj})(M_{ii}+M_{jj})}{M_{ii}^{2}M_{jj}^{2}}E_{ij}.
\end{align*}
Since
\begin{equation*}
\Tr dh(D)(E_{ij})=-4\delta_{ij}\frac{M_{iii}}{M_{ii}^{3}}
\end{equation*}
we have the covariant derivative of the normal vector field
\begin{align*}
&dN(D)(F_{ij})=\delta_{ij}\frac{4}{c(D)^{3}}\frac{M_{iii}}{M_{ii}^{3}}(f'(D))^{-2}
  -\frac{1}{c(D)}\frac{(M_{iij}+M_{ijj})(M_{ii}+M_{jj})}{M_{ii}^{2}M_{jj}^{2}}F_{ij}\\
&dN(D)(H_{ij})=-\frac{1}{c(D)}\frac{(M_{iij}+M_{ijj})(M_{ii}+M_{jj})}{M_{ii}^{2}M_{jj}^{2}}H_{ij}.
\end{align*}
This is the first term in the Equation (\ref{eq:Gamma(D)(X)(N)}).
To get the second term it is enough to substitute the Equation of the normal vector field
  (\ref{eq:N(D) normal vectorfield}) into the explicit formulas for the Christoffel symbol, into the
  Equation (\ref{eq:Gamma(D)(F)(F)}).
\begin{align*}
&\Gamma(D)(F_{ij})(N(D))=\frac{1}{c(D)}\left(\frac{M_{ijj}}{M_{jj}^{2}M_{ij}}
   +\frac{M_{iij}}{M_{ii}^{2}M_{ij}}\right)F_{ij}\\
&\Gamma(D)(H_{ij})(N(D))=\frac{1}{c(D)}\left(\frac{M_{ijj}}{M_{jj}^{2}M_{ij}}
   +\frac{M_{iij}}{M_{ii}^{2}M_{ij}}\right)H_{ij}
\end{align*}
These formulas can be rewritten as
\begin{align*}
&\Gamma(D)(F_{ij})(N)=\delta_{ij}\frac{4}{c(D)^{3}}\frac{M_{iii}}{M_{ii}^{3}}(f'(D))^{-2}
  +\frac{1}{c(D)}\rho_{ij}F_{ij}\\
&\Gamma(D)(H_{ij})(N)=\frac{1}{c(D)}\rho_{ij}H_{ij},\nonumber
\end{align*}
where
\begin{equation*}
\rho_{ij}=\frac{M_{ijj}}{M_{jj}^{2}M_{ij}}+\frac{M_{iij}}{M_{ii}^{2}M_{ij}}
      -\frac{(M_{iij}+M_{ijj})(M_{ii}+M_{jj})}{M_{ii}^{2}M_{jj}^{2}}.
\end{equation*}
The functions $M$ and $\rho$ can be expressed in terms of eigenvalues of the state $D$ and
  the function $f$.
\begin{align*}
M_{ijj}&=\left\lbrace\begin{array}{ll}
  \dfrac{f(\la_{i})-f(\la_{j})}{(\la_{i}-\la_{j})^{2}}+\dfrac{f'(\la_{j})}{\la_{j}-\la_{i}}
             &\quad\mbox{if}\ \la_{i}\neq\la_{j}\\
  \dfrac{1}{2}f''(\la_{i}) &\quad\mbox{if}\ \la_{i}=\la_{j}\end{array}\right.\\
\rho_{ij}&=\left\lbrace\begin{array}{ll}
  -\dfrac{1}{f'(\la_{i})f'(\la_{j})} \dfrac{f'(\la_{i})-f'(\la_{j})}{f(\la_{i})-f(\la_{j})}
  &\quad\mbox{if}\ \la_{i}\neq\la_{j}\\[1em]
  -\dfrac{f''(\la_{i})}{f'(\la_{i})^{3}} &\quad\mbox{if}\ \la_{i}=\la_{j}\end{array}\right.
\end{align*}

After computing the terms in Equation (\ref{eq:Gamma(D)(X)(N)}) and
  substituting in into Equation (\ref{eq:notimportant4}) we have the function $S$.
\begin{align*}
&\mbox{If}\ 1\leq i<j\leq n, 1\leq k<l\leq n:
  &&\hskip-1em\left\lbrace\begin{array}{l} S(D)(H_{ij},H_{kl})=-\frac{2}{c(D)}\delta_{ik}\delta_{jl}\rho_{ij}M_{ij}^{2} \\
                                 S(D)(F_{ij},F_{kl})=-\frac{2}{c(D)}\delta_{ik}\delta_{jl}\rho_{ij}M_{ij}^{2}\\
                                 S(D)(H_{ij},F_{kl})=0.\end{array}\right.\\
&\mbox{If}\ 1\leq i<j\leq n, 1\leq k\leq n:
  &&\hskip-1em\left\lbrace\begin{array}{l} S(D)(H_{ij},F_{kk})=S(D)(F_{kk},H_{ij})=0\\
                                 S(D)(F_{ij},F_{kk})=S(D)(F_{kk},F_{ij})=0.\end{array}\right.\\
&\mbox{If}\ 1\leq i\leq n,   1\leq k\leq n:
  &&\hskip-1em S(D)(F_{ii},F_{kk})=-\frac{8}{c(D)^{3}}\frac{M_{iii}}{M_{ii}^{3}}+\frac{8}{c(D)}\delta_{ik}\frac{M_{iii}}{M_{ii}}.
\end{align*}

The basis of the scalar curvature computation is Equation (\ref{eq:scal main equation}), where
  summation runs on an orthonormal basis of the tangent space of the submanifold.
Fortunately it is no matter if we add the normal vector field to this summation or not,
  as in the classical case, its summand is 0 since
\begin{align*}
S(D)(F_{ij},N(D))&=0,\\
S(D)(H_{ij},N(D))&=0 \\
S(D)(F_{ii},N(D))&=\frac{1}{2c(D)}\sum_{k=1}^{n}\frac{S(D)(F_{ii},F_{kk})}{M_{kk}^{2}}\\
 &=\frac{4}{c(D)^{2}}\frac{M_{iii}}{M_{ii}^{3}}
  \left\lbrack\sum_{k=1}^{n}\left(\frac{-1}{c(D)^{2}M_{kk}^{2}}\right)+1 \right\rbrack=0,\\
S(D)(N(D),N(D))&=\frac{1}{2c(D)}\sum_{k=1}^{n}\frac{1}{M_{kk}^{2}}S(D)(F_{kk},N(D))=0.
\end{align*}

It means that at a given point $D\in\MN$ for an orthonormal basis $(A_{t})_{t\in I}$ in
  $\Ta_{D}\Tilde{\MN}$ the scalar curvature is
\begin{equation*}
\Scal(D)=\sum_{t\in I, s\in I}S(A_{s},A_{s})S(A_{t},A_{t})-S(A_{t},A_{s})S(A_{s},A_{t}).
\end{equation*}

At a point $D\in\MN$ the set of matrices
\begin{equation*}
       \left\lbrace\frac{1}{2M_{ii}}F_{ii}\right\rbrace_{1\leq i\leq n}
\bigcup\left\lbrace\frac{1}{\sqrt{2}M_{ij}}F_{ij}\right\rbrace_{1\leq i<j\leq n}
\bigcup\left\lbrace\frac{1}{\sqrt{2}M_{ij}}H_{ij}\right\rbrace_{1\leq i<j\leq n}
\end{equation*}
form an orthonormal basis in $\Ta_{D}\Tilde{\MN}$ in the case of complex matrices.
It means that we have three kinds of basis elements: diagonal, off-diagonal real and off-diagonal complex ones.

First we consider the case when both $A_{t}$ and $A_{s}$ are diagonal.
\begin{align*}
x_{1}=&\sum_{i,k=1}^{n}\left\lbrack
                       S(D)\left(\frac{1}{2M_{ii}}F_{ii},\frac{1}{2M_{ii}}F_{ii} \right)
                       S(D)\left(\frac{1}{2M_{kk}}F_{kk},\frac{1}{2M_{kk}}F_{kk} \right)\right.\\
&\left.                -S(D)\left(\frac{1}{2M_{ii}}F_{ii},\frac{1}{2M_{kk}}F_{kk} \right)
                       S(D)\left(\frac{1}{2M_{kk}}F_{kk},\frac{1}{2M_{ii}}F_{ii} \right)\right\rbrack\\
&=\frac{1}{16}\sum_{i,k=1}^{n}\frac{64M_{iii}M_{kkk}}{c(D)^{2}M_{ii}^{3}M_{kk}^{3}}
  \left\lbrack\left(-\frac{1}{c(D)^{2}}\frac{1}{M_{ii}^{2}}+1 \right)
              \left(-\frac{1}{c(D)^{2}}\frac{1}{M_{kk}^{2}}+1 \right)\right.\\
&\left.       -\left(-\frac{1}{c(D)^{2}}\frac{1}{M_{ii}^{2}}+\delta_{ik} \right)
              \left(-\frac{1}{c(D)^{2}}\frac{1}{M_{kk}^{2}}+\delta_{ik} \right) \right\rbrack\\
&=\frac{4}{c(D)^{2}}\sum_{\btop{i,k=1}{i\neq k}}^{n}\frac{M_{iii}M_{kkk}}{M_{ii}^{3}M_{kk}^{3}}
  \left(1-\frac{1}{c(D)^{2}M_{ii}^{2}}-\frac{1}{c(D)^{2}M_{kk}^{2}}\right)
\end{align*}

\vfill\eject\newpage

If $A_{t}$ is diagonal and $A_{s}$ is off-diagonal real matrices then we have
\begin{align*}
x_{2}=&\sum_{\btop{1\leq i<j\leq n}{1\leq k\leq n}}\left\lbrack
                       S(D)\left(\frac{1}{\sqrt{2}M_{ij}}F_{ij},\frac{1}{\sqrt{2}M_{ij}}F_{ij} \right)
                       S(D)\left(\frac{1}{2M_{kk}}F_{kk},\frac{1}{2M_{kk}}F_{kk} \right)\right.\\
&\left.                -S(D)\left(\frac{1}{2M_{kk}}F_{kk},\frac{1}{\sqrt{2}M_{ij}}F_{ij} \right)
                       S(D)\left(\frac{1}{\sqrt{2}M_{ij}}F_{ij},\frac{1}{2M_{kk}}F_{kk}\right)\right\rbrack\\
&=\frac{1}{8}\sum_{k=1}^{n}\frac{16M_{kkk}}{c(D)^{2}M_{kk}^{3}}
              \left(-\frac{1}{c(D)^{2}}\frac{1}{M_{kk}^{2}}+1 \right)
              \left(\sum_{1\leq i<j\leq n}-\rho_{ij}\right)\\
&=-\frac{2}{c(D)^{2}}\left\lbrack\sum_{k=1}^{n}\frac{M_{kkk}}{M_{kk}^{3}}
                   \left(1-\frac{1}{c(D)^{2}M_{kk}^{2}}\right)\right\rbrack
             \left(\sum_{1\leq i<j\leq n}\rho_{ij}\right).
\end{align*}

If $A_{t}$ is diagonal and $A_{s}$ is off-diagonal complex matrix then because of the
  equation $S(F_{ij},F_{ji})=S(H_{ij},H_{ij})$ the summation will be equal to $x_{2}$.
If $A_{t}$ is off-diagonal real matrix and $A_{s}$ is diagonal one, then because the
  summation is symmetric the result will be $x_{2}$ again.

If both $A_{t}$ and $A_{s}$ are off-diagonal real matrices then the summation is the following.
\begin{align*}
x_{3}=&\sum_{\btop{1\leq i<j\leq n}{1\leq k<l\leq n}}\left\lbrack
                       S(D)\left(\frac{1}{\sqrt{2}M_{ij}}F_{ij},\frac{1}{\sqrt{2}M_{ij}}F_{ij} \right)
                       S(D)\left(\frac{1}{\sqrt{2}M_{kl}}F_{kl},\frac{1}{\sqrt{2}M_{kl}}F_{kl} \right)\right.\\
&\left.                -S(D)\left(\frac{1}{\sqrt{2}M_{kl}}F_{kl},\frac{1}{\sqrt{2}M_{ij}}F_{ij} \right)
                       S(D)\left(\frac{1}{\sqrt{2}M_{ij}}F_{ij},\frac{1}{\sqrt{2}M_{kl}}F_{kl} \right)\right\rbrack\\
&=\frac{1}{4}\sum_{\btop{1\leq i<j\leq n}{1\leq k<l\leq n}}\frac{1}{M_{ij}^{2}M_{kl}^{2}}
              \left(\frac{2}{c(D)}\rho_{ij}M_{ij}^{2}\frac{2}{c(D)}\rho_{kl}M_{kl}^{2}
                    -\delta_{ik}\delta_{jl}\frac{4}{c(D)^{2}}\rho_{ij}^{2}M_{ij}^{4} \right)\\
&=\frac{1}{c(D)^{2}}\sum_{\btop{1\leq i<j\leq n}{1\leq k<l\leq n}}
                      (\rho_{ij}\rho_{kl}-\delta_{ik}\delta_{jl}\rho_{ij}^{2})\\
&= \frac{1}{c(D)^{2}}\left(\sum_{1\leq i<j\leq n}\rho_{ij}\right)^{2}
  -\frac{1}{c(D)^{2}}\sum_{1\leq i<j\leq n}\rho_{ij}^{2}
\end{align*}

If both $A_{t}$ and $A_{s}$ are off-diagonal matrices, but $A_{t}$ is real and $A_{s}$ is a complex one then
\begin{align*}
x_{4}=&\sum_{\btop{1\leq i<j\leq n}{1\leq k<l\leq n}}\left\lbrack
                       S(D)\left(\frac{1}{\sqrt{2}M_{ij}}H_{ij},\frac{1}{\sqrt{2}M_{ij}}H_{ij} \right)
                       S(D)\left(\frac{1}{\sqrt{2}M_{kl}}F_{kl},\frac{1}{\sqrt{2}M_{kl}}F_{kl} \right)\right.\\
&\left.                -S(D)\left(\frac{1}{\sqrt{2}M_{kl}}F_{kl},\frac{1}{\sqrt{2}M_{ij}}H_{ij} \right)
                       S(D)\left(\frac{1}{\sqrt{2}M_{ij}}H_{ij},\frac{1}{\sqrt{2}M_{kl}}F_{kl} \right)\right\rbrack\\
&=\frac{1}{4}\sum_{\btop{1\leq i<j\leq n}{1\leq k<l\leq n}}\frac{1}{M_{ij}^{2}M_{kl}^{2}}
              \left(\frac{2}{c(D)}\rho_{ij}M_{ij}^{2}\frac{2}{c(D)}\rho_{kl}M_{kl}^{2}\right)
=\frac{1}{c(D)^{2}}\sum_{\btop{1\leq i<j\leq n}{1\leq k<l\leq n}}\rho_{ij}\rho_{kl}.
%=\frac{1}{c(D)^{2}}\left(\sum_{1\leq i<j\leq n}\rho_{ij}\right)^{2} .
\end{align*}

If $A_{t}$ is an off-diagonal complex matrix, and $A_{s}$ is a diagonal one or off-diagonal real one
  then result of the summation is equal to $x_{2}$ and $x_{4}$.
Finally, if both $A_{t}$ and $A_{s}$ are off-diagonal complex matrices then because of the
  equation $S(F_{ij},F_{ji})=S(H_{ij},H_{ij})$ the summation is equal to $x_{3}$.

The classical state space $\Pn$ corresponds to the case, when $\MN$ consists of only diagonal elements,
  that is, the scalar curvature of this space is $x_{1}$.
If $\MN$ contains only real elements, then the scalar curvature is $x_{1}+2x_{2}+x_{3}$,
  and in the complex case the curvature is $x_{1}+4x_{2}+2x_{3}+2x_{4}$.
Combining the computations we have the following theorem.

\begin{theorem}
The scalar curvature of the real and complex state space $(\MN,g_{f})$ for an admissible function
  $f$ at a point $D\in\MN$ with eigenvalues $(\lambda_{i})_{i=1,\dots,n}$ is
\begin{align}\label{eq:scal quantum equation}
&\Scal(D)_{\mathbb{R}}=\frac{4}{c(D)^{4}}\left\lbrack
\sum_{\btop{i,k=1}{i\neq k}}^{n}\frac{M_{iii}M_{kkk}}{M_{ii}^{3}M_{kk}^{3}}
  \left(c(D)^{2}-\frac{1}{M_{ii}^{2}}-\frac{1}{M_{kk}^{2}}\right) \right.\\
&\ \left.-\left( \sum_{k=1}^{n}\frac{M_{kkk}}{M_{kk}^{3}}
                   \left(c(D)^{2}-\frac{1}{M_{kk}^{2}}\right)\right)
           \left(\sum_{1\leq i<j\leq n}\rho_{ij}\right)\right\rbrack
 +\frac{1}{c(D)^{2}}\left(\sum_{1\leq i<j\leq n}\rho_{ij}\right)^{2}\nonumber\\
&\ -\frac{1}{c(D)^{2}}\sum_{1\leq i<j\leq n}\rho_{ij}^{2}\nonumber\\
&\Scal(D)_{\mathbb{C}}=\frac{4}{c(D)^{4}}\left\lbrack
\sum_{\btop{i,k=1}{i\neq k}}^{n}\frac{M_{iii}M_{kkk}}{M_{ii}^{3}M_{kk}^{3}}
  \left(c(D)^{2}-\frac{1}{M_{ii}^{2}}-\frac{1}{M_{kk}^{2}}\right) \right.\nonumber\\
&\ \left.-2\left( \sum_{k=1}^{n}\frac{M_{kkk}}{M_{kk}^{3}}
                   \left(c(D)^{2}-\frac{1}{M_{kk}^{2}}\right)\right)
           \left(\sum_{1\leq i<j\leq n}\rho_{ij}\right)\right\rbrack
 +\frac{4}{c(D)^{2}}\left(\sum_{1\leq i<j\leq n}\rho_{ij}\right)^{2}\nonumber\\
&\ -\frac{2}{c(D)^{2}}\sum_{1\leq i<j\leq n}\rho_{ij}^{2},\nonumber
\end{align}
where
\begin{align*}
& M_{ii}=f'(\la_{i}),\quad M_{iii}=\frac{f''(\la_{i})}{2},
  \quad c(D)=\sqrt{\sum_{k=1}^{n}\frac{1}{f'(\la_{k})^{2}}},\\
&\rho_{ij}=\left\lbrace\begin{array}{ll}
  -\dfrac{1}{f'(\la_{i})f'(\la_{j})} \dfrac{f'(\la_{i})-f'(\la_{j})}{f(\la_{i})-f(\la_{j})}
  &\quad\mbox{if}\ \la_{i}\neq\la_{j}\\[1em]
  -\dfrac{f''(\la_{i})}{f'(\la_{i})^{3}} &\quad\mbox{if}\ \la_{i}=\la_{j}.\end{array}\right.
\end{align*}
\end{theorem}

We can test the theorem  in three different cases.
As it was mentioned, if we restrict ourselves to the functions of the form
  $f(x)=\frac{2}{1-\alpha}x^{\frac{1-\alpha}{2}}$ and
  to diagonal matrices, then we get back the scalar curvature of the classical $\alpha$-geometry.
Let $D\in\MN$ be a diagonal state and denote by $\vt$ the diagonal elements of $D$.
In this case $c(D)=c(\vt)$ and the term $x_{1}$ is
\begin{align*}
x_{1}&=\frac{4}{c(D)^{2}}\sum_{\btop{i,k=1}{i\neq k}}^{n}\frac{M_{iii}M_{kkk}}{M_{ii}^{3}M_{kk}^{3}}
  \left(1-\frac{1}{c(D)^{2}M_{ii}^{2}}-\frac{1}{c(D)^{2}M_{kk}^{2}}\right)
\end{align*}
\begin{align*}
  &=\frac{4}{c(\vt)^{2}}\sum_{\btop{i,k=1}{i\neq k}}^{n}
   \frac{f''(\vt_{i})f''(\vt_{k})}{4 f'(\vt_{i})^{3} f'(\vt_{k})^{3} }
  \left(1-\frac{1}{c(\vt)^{2} f'(\vt_{i})^{2}}-\frac{1}{c(\vt)^{2} f'(\vt_{k})^{2}}\right)\\
  &=\frac{1}{c(\vt)^{2}}\sum_{\btop{i,k=1}{i\neq k}}^{n}\left(\frac{1+\alpha}{2}\right)^{2}
  \left(1-\frac{\vt_{i}^{\alpha+1}+\vt_{k}^{\alpha+1}}{c(\vt)^2} \right)
\end{align*}
which is equal to Equation (\ref{eq:scal classical equation}).
If the function is $f(x)=\log x$ then
\begin{equation}\label{eq:scal classical equation log}
x_{1}=\frac{1}{c(\vt)^{2}}\sum_{\btop{i,k=1}{i\neq k}}^{n}\vt_{i}\vt_{k}
  \left(1-\frac{\vt_{i}^{2}+\vt_{k}^{2}}{c(\vt)^2} \right)
\end{equation}
which is equal to Equation (\ref{eq:scal classical equation}) if $\alpha=1$.

If we consider the full real or complex state space $\MN$ and the function is $f(x)=2\sqrt{x}$ then the
  pull-back metric is the Wigner--Yanase metric \cite{WigYan}, as this has been proved by Gibilisco and
  Isola \cite{GibIso2}.
In this case we map the state space to the surface of an Euclidean ball with radius $R=2$.
For a given state $D\in\MN$ with eigenvalues $(\lambda_{1},\dots,\lambda_{n})$ we have $c(D)=1$ and
  $\rho_{ij}=\frac{1}{2}$.
The terms $x$ in the scalar curvature formula are the following.
\begin{align*}
x_{1}&=4\sum_{\btop{i,k=1}{i\neq k}}^{n}
        \frac{f''(\vt_{i})f''(\vt_{k})}{4 f'(\vt_{i})^{3} f'(\vt_{k})^{3} }
        \left(1-\frac{1}{f'(\vt_{i})^{2}}-\frac{1}{f'(\vt_{k})^{2}}\right)\\
     &=\frac{1}{4}\sum_{\btop{i,k=1}{i\neq k}}^{n}(1-\vt_{i}-\vt_{k})=\frac{n^{2}-3n+2}{4}\\
x_{2}&=-2\left(\sum_{k=1}^{n}-\frac{1}{4}(1-\lambda_{k}) \right)
     \left(\sum_{1\leq i<j\leq n}\frac{1}{2}\right)=\frac{n(n-1)^2}{8}\\
x_{3}&=\left(\sum_{1\leq i<j\leq n}\frac{1}{2}\right)^{2}-
       \sum_{1\leq i<j\leq n}\frac{1}{4}=\frac{n^{2}(n-1)^{2}}{16}-\frac{n(n-1)}{8}\\
x_{4}&=\left(\sum_{1\leq i<j\leq n}\frac{1}{2}\right)^{2}=\frac{n^{2}(n-1)^{2}}{16}
\end{align*}
The scalar curvature of the real and complex state spaces are
\begin{align*}
\Scal_{\mathbb{R}}(D)&=x_{1}+2x_{2}+x_{3}=\frac{d_{\mathbb{R}}(d_{\mathbb{R}} -1)}{R^{2}}\\
\Scal_{\mathbb{C}}(D)&=x_{1}+4x_{2}+2x_{3}+2x_{4}=\frac{d_{\mathbb{C}}(d_{\mathbb{C}} -1)}{R^{2}}
\end{align*}
which are just the well-known scalar curvatures of the Euclidean spheres in dimensions $d_{\mathbb{R}}$
  and $d_{\mathbb{C}}$ with radius $R$.

Finally if we use the $f(x)=x$ function, then we map the state space into the flat Euclidean space,
  so the scalar curvature is $0$, and we have the same result from Equation
  (\ref{eq:scal quantum equation}).

The scalar curvature formula can be simplified, if we consider only $2\times 2$ density matrices.
In this case $x_{1}=0$ and $x_{3}=0$ and the $x_{2}$, $x_{4}$ terms are given in the following
  Corollary.

\begin{corollary}\label{co:scalar curvature of M2}
The scalar curvature of the real and complex state space $(\Mtwo,g_{f})$ for an admissible function
  $f$ at a point $D\in\Mtwo$ with eigenvalues $\la_{1}$ and $\la_{2}$ are
\begin{equation*}
\Scal(D)_{\mathbb{R}}=2x_{2} \qquad
\Scal(D)_{\mathbb{C}}=4x_{2}+2x_{4},
\end{equation*}
where
\begin{align*}
&x_{2}=\frac{f'(\la_{1})f'(\la_{2})}{\left(f'(\la_{1})^{2}+f'(\la_{2})^{2} \right)^{2}}
  \left(\frac{f''(\la_{1})}{f'(\la_{1})}+\frac{f''(\la_{2})}{f'(\la_{2})} \right)
  \frac{f'(\la_{1})-f'(\la_{2})}{f(\la_{1})-f(\la_{2})}\\
&x_{4}=\frac{1}{f'(\la_{1})^{2}+f'(\la_{2})^{2}}
  \left(\frac{f'(\la_{1})-f'(\la_{2})}{f(\la_{1})-f(\la_{2})}\right)^{2}.
\end{align*}
\end{corollary}

\section{Monotone scalar curvatures}

Now we can formalize Petz's conjecture \cite{Pet2}:
The scalar curvature of the space $(\MN,g_{\KM})$ is monotonously decreasing with respect the
  majorization relation, that is for every states $D_{1},D_{2}\in\MN$ if $D_{1}\prec D_{2}$ then
  $\Scal(D_{1})\geq\Scal(D_{2})$.

The corresponding conjecture to the case of $\alpha$-geometries is due to Gibilisco and Isola \cite{GibIso1}:
On the spaces $(\Pn,g_{\alpha})$ and $(\MN,g_{\alpha})$ the scalar curvature is monotonously increasing,
 with respect to the majorization relation if $\alpha\in\left\rbrack -1,0\right\lbrack$ and
 it is monotonously decreasing if $\alpha\in\left\rbrack 0,1\right\lbrack$.
They proved a similar statement for the curvature of the space $(\Ptwo,g_{\alpha})$.

A linear map $T$ on $\mathbb{R}^{n}$ is a $T$-transform if there exists $0\leq t\leq 1$ and indices $k,l$ such that
$T(x_{1},\dots,x_{n})$ is equal to
\begin{equation*}
(x_{1},\dots,x_{k-1},tx_{k}+(t-1)x_{l},x_{k+1},\dots,x_{l-1},(1-t)x_{k}+tx_{l},x_{l+1},\dots,x_{n}).
\end{equation*}
For every $a\in\Pn$ and for every $T$ transform $T(a)\prec a$.
For given $a,b\in\Pn$ if $a\prec b$, then we can go continuously from $a$ to $b$ using only
  $T$-transformations \cite{Bha}.

\begin{theorem}
Assume that we have $a,b\in\Pn$ with decreasingly ordered elements $(a_{1},\dots,a_{n})$ and $(b_{1},\dots,b_{n})$.
The following statements are equivalent.
\begin{enumerate}
\item The distribution $a$ is more mixed than $b$.
\item One can find a sequence $(c_{z})_{z=1,\dots,d}$ between them such that for all $z=1,\dots,d$: $c_{z}\in\Pn$,
\begin{equation*}
a=c_{1}\prec c_{2}\prec\dots\prec c_{d}=b
\end{equation*}
  holds and the set of values of $c_{z}$ and $c_{z-1}$ is the same except two elements.
\item The set $(a_{1},\dots,a_{n})$ can be obtained from $(b_{1},\dots,b_{n})$
      by a finite number of T-transforms.
\end{enumerate}
\end{theorem}

According to this Theorem in order to prove the monotonicity of the scalar curvature with respect
  to the majorization, it is enough to consider those distributions which have only two different elements.
For example if we consider the space $(\Pthree,g_{\alpha})$ and we combine the previous Theorem with
  Corollary (\ref{co:scalar curvature of P3}) we have the following simplification for the conjecture.

\begin{corollary}
To prove Gibilisco's and Isola's Conjecture for the space $(\Pthree,g_{\alpha})$ it is enough to show that for
  every distribution $(a_{1},a_{2},a_{3})\in\Pthree$ if $a_{1}>a_{2}$ then the function
\begin{equation*}
 \left\lbrack 0,\frac{a_{1}-a_{2}}{2} \right\rbrack\to\mathbb{R}\qquad x\mapsto
\frac{(a_{1}-x)^{\alpha}(a_{2}+x)^{\alpha}}{\bigl( (a_{1}-x)^{\alpha+1}+(a_{2}+x)^{\alpha+1}+a_{3}^{\alpha+1}\bigr)^{2}}
\end{equation*}
  is decreasing if $\alpha\in\left\rbrack -1,0\right\lbrack$ and increasing if
$\alpha\in\left\rbrack 0,1\right\lbrack$.
\end{corollary}

If we consider the space $(\Mtwo,g_{\alpha})$ we can compute the scalar curvature according to
  Corollary (\ref{co:scalar curvature of M2}).
We write the eigenvalues of a state $D\in\Mtwo$ as $\dfrac{r+1}{2}$ and $\dfrac{r-1}{2}$, where $r$ is
  the interval $\left]0,1\right[$.
Using this parameter, for states $D_{1},D_{2}\in\Mtwo$ the relation $D_{1}\prec D_{2}$ holds if and
  only if $r_{1}\leq r_{2}$.
Numerically we computed the scalar curvature of the state space $(\Mtwo,g_{\alpha})$ using Maple.
The scalar curvature of the real state space can be seen on the following graphs.
\begin{center}
\resizebox{50mm}{!}{\rotatebox{0}{\includegraphics{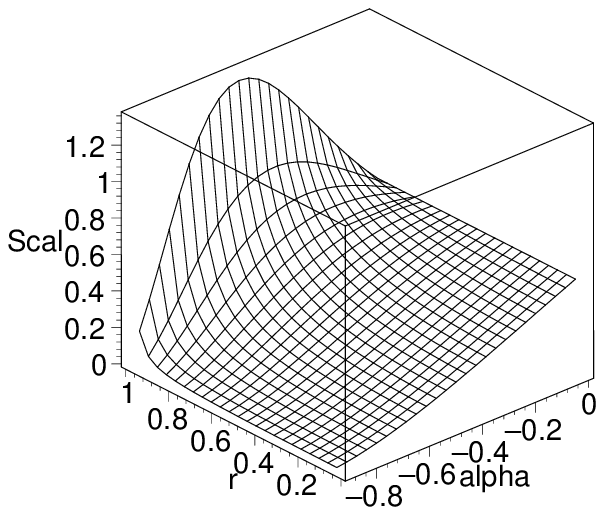}}}
\hskip1cm
\resizebox{50mm}{!}{\rotatebox{0}{\includegraphics{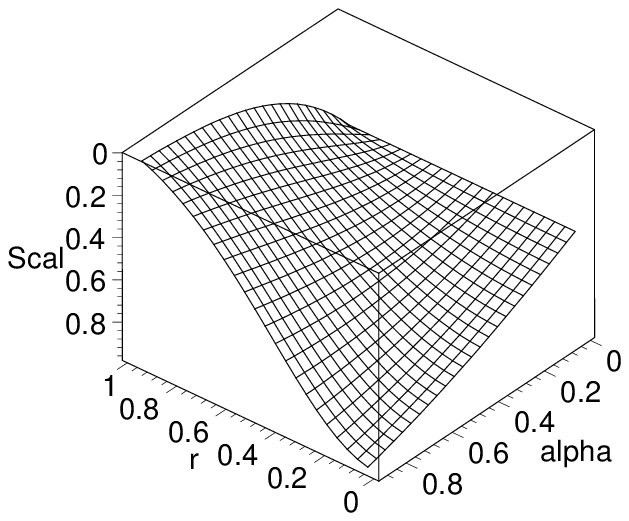}}}
\end{center}
It seems that the scalar curvature is increasing with respect to the majorization if $\alpha\in\left]-1,0\right[$
  and decreasing for parameters $\alpha\in\left]0,1\right[$.

The following graphs are about the scalar curvature of the complex state space $(\Mtwo,g_{\alpha})$.
\begin{center}
\resizebox{50mm}{!}{\rotatebox{0}{\includegraphics{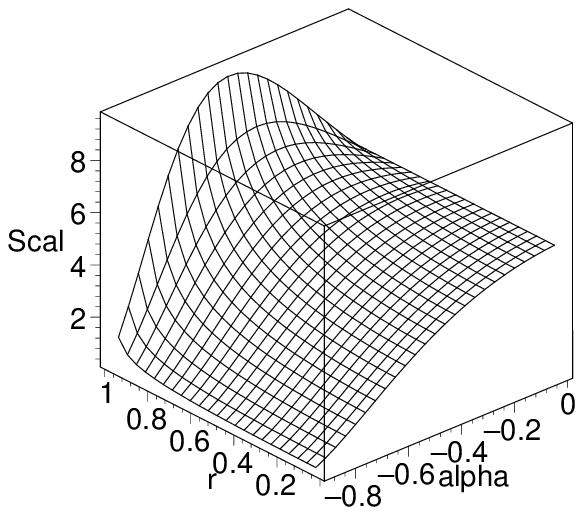}}}
\hskip1cm
\resizebox{50mm}{!}{\rotatebox{0}{\includegraphics{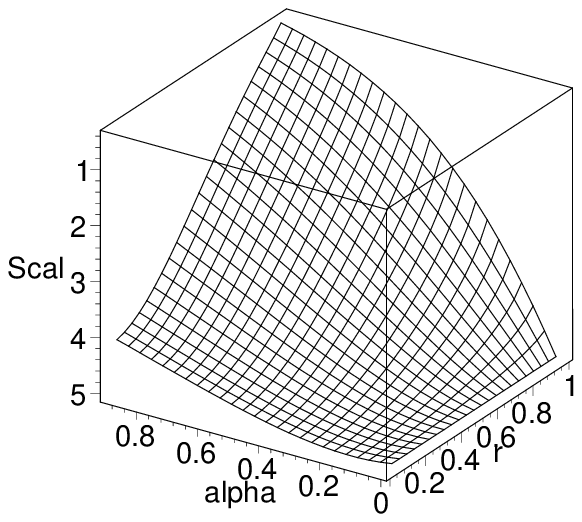}}}
\end{center}
We can check again that the foreseen properties of the scalar curvature function seems to be true.

\section*{Acknowledgements}
This work was supported by Hungarian Scientific Research Fund (OTKA)
  contract T046599, T43242, TS049835,
  and EU Network ''QP-Applications'' contract number HPRN--CT--2002--00729.

\end{document}